\documentclass[draft]{agujournal}
\draftfalse

\usepackage{amsmath,amssymb,graphicx}

\journalname{JGR-Planets}

\begin{document}

%
%


\title{Quantitative estimates of chemical disequilibrium in Titan's atmosphere}

%
%




\authors{B. F. M. Intoy\affil{1} and J. W. Halley\affil{1}}


\affiliation{1}{School of Physics and Astronomy, University of Minnesota, Minneapolis, Minnesota, USA}




\correspondingauthor{J. W. Halley}{halle001@umn.edu}




\begin{keypoints}
\item  A nearly exact analysis of  a statistical mechanical model for estimating 
the degree of disequilibrium in Titan's atmosphere is  shown to agree with a previous 
approximate analysis.
\item  The estimated measures of disequilibrium of Titan's atmosphere lie between those
of biological systems and some engineered polymer systems. 
\item  Some new features of the mathematical treatment of the model are described. 
\end{keypoints}

%
%


\begin{abstract}
We apply previously introduced measures of chemical disequilibrium to Cassini
mass spectroscopy data on the atmosphere of Titan.   In the analysis presented here, we use
an improved description, avoiding the meanfield approximation in previous work.
The results of the analysis are nearly exactly  the same as those found earlier 
and confirm that, with respect to the measures used, Titan's atmosphere lies between
living and many nonliving systems. Some details of the mathematical analysis, which 
appear to be new, are included. 
\end{abstract}

\section{Introduction}


The atmosphere of Titan has long been speculated 
to have an atmosphere similar to that of early earth which might serve as a 
model for prebiotic evolution\citep{clarke1997chemical,trainer2006organic}.  
In the course of a recent study of data 
on that atmosphere from
the NASA Cassini-Huygens mission to Saturn, we formulated \citep{2018pub} an approximate  model for 
estimating how far that atmosphere is from  
chemical equilibrium. The model took  the form of a ferrimagnetic 
Ising model for each of multiple linear chain molecules. In \citet{2018pub} we made 
an uncontrolled approximation, a kind of mean field theory, to determine the equilibrium
states of the model in analyzing the Titan data.  Here we report a more exact analysis
which does not make that approximation.

Titan has a dense atmosphere made  mostly of nitrogen.
Methane gas is present, with concentrations of about 2 atomic \% \citep{waite2007process},
which precipitates and cycles out of the atmosphere \citep{lunine2008methane}.
as well as larger molecules up to  10,000 atomic mass units which were
detected in the
atmosphere on the mass spectrometer instruments of the Cassini spacecraft 
\citep{waite2007process}. Mass spectrometry data are available for the negatively
charged, neutral and positively charged molecules in the atmosphere.  The
most massive detected molecules were negatively charged. 
The model presented here is intended to model the equilibrium distributions 
of those larger  molecules which are believed  to be mainly composed of 
nitrogen, carbon, and hydrogen.  Although it is  possible
that these large molecules could have complex structures,
we have assumed in the model that they are linear chains and we used 
a 'united atom' model in which the hydrogen entities are not treated explicitly.

An uncontrolled approximation for the partition function in the equilibrium description of the model reported here was 
used in \citet{2018pub} to estimate the degree to which the 
atmosphere of Titan is out of local chemical equilibrium. and out
of chemical equilibrium  with an external thermal bath at the 
reported ambient temperature of that atmosphere. 
Here we report details of an exact solution for the equilibrium partition 
function of the model.
In the last section of the paper, we report results of the same 
disequilibrium calculations described in \citet{2018pub} using 
the more exact equilibrium description given here.

 In the next section, we describe the single chain model, its
 extension to many chains, the way in which spatial dilution was
taken into account and the Gibbs limit of large negative chemical 
potential which we will use in the analysis . In the third section
we describe calculations of 
disequilibrium of Titan's atmosphere like those reported in
our previous work \citet{2018pub}
and compare the new results with those of those previous 
approximate calculations.

\section{Description of the  Model}

We consider a collection of linear chain molecules consisting of monomers
of two types, which we regard in the application as being 'united atom' descriptions 
of  carbon and nitrogen plus some hydrogen atoms. Denoting the two entities as
C and N, and motivated by the C-C C-N and N-N bond energies reported from 
first  principles calculations in Table \ref{tab:cnbond} we choose a model in which 
those bond energies obey the relations $\Delta_{CC}=\Delta_{CN}=\Delta_1$
and $\Delta_{NN}=\Delta_2$. (In the numerical calculations reported in section 
IV we used $\Delta_1=325  kJ/mol $ and $\Delta_2=160 kJ/mol$. )
The relative concentration of monomers of the two types, which is known 
experimentally \citep{crary2009heavy} , is controlled in the model with 
a magnetic field-like parameter $h$.


\begin{table}[h]
 \caption{The average bond energies for carbon and nitrogen \citep{zumdahl2007chemistry}.}
 \label{tab:cnbond}
 \begin{tabular}{ c  c }
  \hline
  Bond & Average Bond Energy (kJ/mol) \\
  \hline
  C-C & 347 \\ 
  C-N & 305 \\
  N-N & 160 \\
  \hline
 \end{tabular}
\end{table}

With those assumptions and that parametrization, the model for a single 
chain with number of monomers $L$ takes the form of a 
ferromagnetic one dimensional Ising model
\begin{linenomath*}
\begin{equation} \label{eq:singhamil}
H(\boldsymbol{\sigma})=-\sum_{i=1}^{L-1} J(\sigma_i,\sigma_{i+1}) \, \sigma_i \sigma_{i+1}-h \sum_{i=1}^{L} \sigma_i
\end{equation}
\end{linenomath*}
where $\sigma_i$ takes the values $\{+1,-1\}$  referring respectively to carbon and  nitrogen monomers.
With the parametrization of the bond energies described above, the interaction matrix 
 $J(\sigma_i,\sigma_j)$ takes the form
\begin{linenomath*}
\begin{equation} \label{eq:Jmatrix}
\boldsymbol{J} \equiv
 \begin{pmatrix}
 J(+,+) & J(+,-) \\
 J(-,+) & J(-,-) \\
 \end{pmatrix}
=
 \begin{pmatrix}
 \Delta_1 & -\Delta_1 \\
 -\Delta_1 & \Delta_2 \\
 \end{pmatrix}
\end{equation} 
\end{linenomath*}
where $\Delta_1>\Delta_2>0$.
 $h$ controls the relative concentration of C and N as mentioned above.
The partition function is
\begin{linenomath*}
\begin{equation} \label{eq:ZLfull}
Z_L = \sum_{\boldsymbol{\sigma}} \exp \left [ -\beta H(\boldsymbol{\sigma})  \right ]
 = \sum_{\boldsymbol{\sigma}} \exp \left [ \beta \sum_{i=1}^{L-1} J(\sigma_i,\sigma_{i+1}) \, \sigma_i \sigma_{i+1}+\beta h \sum_{i=1}^{L} \sigma_i \right ].
\end{equation}
\end{linenomath*}
where $\beta^{-1}=k_B T$ with $T$ the absolute temperature and $k_B$ Boltzmann's constant.
Using the transfer matrix method \citep{kramers} the exponential in equation \ref{eq:ZLfull}
is  factored into terms involving only two neighboring monomers:
\begin{linenomath*}
\begin{align} \label{eq:ZLMs}
Z_L = & \sum_{\boldsymbol{\sigma}} \exp \left [\frac{- \beta  h \sigma_1}{2} \right ] \, M(\sigma_1,\sigma_2) \, M(\sigma_2,\sigma_3) \cdots \nonumber \\ 
      & \cdots \, M(\sigma_{L-2},\sigma_{L-1}) \, M(\sigma_{L-1},\sigma_{L}) \, \exp \left [\frac{- \beta  h \sigma_L}{2} \right ],
\end{align}
\end{linenomath*}
where
\begin{linenomath*}
\begin{equation}
M(\sigma_i,\sigma_j) \equiv \exp \left [ \beta J(\sigma_i,\sigma_j) \, \sigma_i \sigma_j + \frac{\beta h}{2}(\sigma_i+\sigma_j) \right ].
\end{equation}
\end{linenomath*}
Written out as a matrix, $\boldsymbol{M}$ has the form:
\begin{linenomath*}
\begin{equation} \label{eq:Mmatrix}
\boldsymbol{M} \equiv
 \begin{pmatrix}
 M(+,+) & M(+,-) \\
 M(-,+) & M(-,-) \\
 \end{pmatrix}
=
 \begin{pmatrix}
 e^{\beta ( \Delta_1 + h)} & e^{ \beta \Delta_1} \\
 e^{ \beta \Delta_1} & e^{\beta ( \Delta_2 - h)} \\
 \end{pmatrix}.
\end{equation} 
\end{linenomath*}
In equation \ref{eq:ZLMs} the summations over $\sigma_2, \sigma_3, \cdots , \sigma_{L-1}$
are matrix multiplications.  The partition function is then
\begin{linenomath*}
\begin{equation} \label{eq:ZLMmat}
Z_L =  \sum_{\sigma_1,\sigma_L} \exp \left [\frac{- \beta  h \sigma_1}{2} \right ] \left [ \boldsymbol{M}^{L-1} \right ]_{\sigma_1,\sigma_L}  \exp \left [\frac{- \beta  h \sigma_L}{2} \right ].
\end{equation}
\end{linenomath*}
Since $\boldsymbol{M}$ is a symmetric matrix there exists a unitary matrix $\boldsymbol{P}$, constructed from  the
 eigenvectors of $\boldsymbol{M}$, such that
$\boldsymbol{M}=\boldsymbol{P}\boldsymbol{D}\boldsymbol{P}^{-1}$, where $\boldsymbol{D}$ is a diagonal
matrix containing the eigenvalues of $\boldsymbol{M}$.
Solving for the eigenvalues ($\lambda_\pm$) and eigenvectors ($\boldsymbol{x}_\pm$) yields:
\begin{linenomath*}
\begin{gather}
\lambda_\pm = \frac{1}{2} \left [ (ac+bc^{-1}) \pm \sqrt{4a^2+(ac-bc^{-1})^2} \right ] \\
\boldsymbol{x}_\pm = \frac{1}{\sqrt{a^2+(ac-\lambda_\pm)^2}}
  \begin{pmatrix}
   -a \\
   ac - \lambda_\pm
  \end{pmatrix}
\end{gather}
\end{linenomath*}
where $a$, $b$, and $c$ are defined as
\begin{linenomath*}
\begin{equation} \label{eq:abcdef}
a \equiv \exp (\beta \Delta_1) \ , \ b \equiv \exp (\beta \Delta_2) \ , \ c \equiv \exp (\beta h)
\end{equation}
\end{linenomath*}
.Note that $|\lambda_+|>|\lambda_-|$.
  The matrix multiplication in equation \ref{eq:ZLMmat}
becomes
 $\boldsymbol{M}^{L-1}=(\boldsymbol{P}\boldsymbol{D}\boldsymbol{P}^{-1})^{L-1}=\boldsymbol{P}\boldsymbol{D}^{L-1}\boldsymbol{P}^{-1}$.
 Where $\boldsymbol{P}=(\boldsymbol{x}_+,\boldsymbol{x}_-)$ and $\boldsymbol{P}^{-1}=\boldsymbol{P}^T$ since $\boldsymbol{P}$ is unitary.
Substituting into equation \ref{eq:ZLMmat} and  summing over $\sigma_1$ and $\sigma_L$ gives:
\begin{linenomath*}
 \begin{align}
  Z_L & = \sum_{\sigma_1,\sigma_L} \exp \left [\frac{- \beta  h \sigma_1}{2} \right ] \left [ \boldsymbol{M}^{L-1} \right ]_{\sigma_1,\sigma_L}  \exp \left [\frac{- \beta  h \sigma_L}{2} \right ] \\
      & = \sum_{\sigma_1,\sigma_L} \exp \left [\frac{- \beta  h \sigma_1}{2} \right ] \left [ \boldsymbol{P}\boldsymbol{D}^{L-1}\boldsymbol{P}^{-1} \right ]_{\sigma_1,\sigma_L}  \exp \left [\frac{- \beta  h \sigma_L}{2} \right ] \\
      & = \sum_{\sigma_1,\sigma_L} \exp \left [\frac{- \beta  h \sigma_1}{2} \right ] \Bigg [ \boldsymbol{P}
            \begin{pmatrix}
              \lambda_+^{L-1} & 0 \\
              0 & \lambda_-^{L-1} \\
            \end{pmatrix}
          \boldsymbol{P}^T \Bigg ]_{\sigma_1,\sigma_L}  \exp \left [\frac{- \beta  h \sigma_L}{2} \right ] \\
      & = \frac{1}{c} \left ( \frac{\lambda_+^{L+1}}{a^2+(ac-\lambda_+)^2} + \frac{\lambda_-^{L+1}}{a^2+(ac-\lambda_-)^2} \right )  \label{eq:ZLabcraw}
 \end{align}
\end{linenomath*}

 \begin{table}
 \caption{
    $Z_L$ for small values of $L$ using $\beta, h$ notation.
     \label{tab:ZLDeltah} }
 \begin{tabular}{c c}
 \hline
 $L$ & $Z_L(h,\beta)$ \\
 \hline
 2 & $e^{\beta(\Delta_1+2h)}+2e^{\beta(\Delta_1)}+e^{\beta(\Delta_2-2h)}$ \\
 3 & $e^{\beta(2\Delta_1+3h)}+3e^{\beta(2\Delta_1+h)}+e^{\beta(2\Delta_1-h)}+2e^{\beta(\Delta_1+\Delta_2-h)}+e^{\beta(2\Delta_2-3h)}$ \\
 4 & $e^{\beta(3\Delta_1+4h)}+4e^{\beta(3\Delta_1+2h)}+3e^{\beta (3\Delta_1)}+3e^{\beta(2\Delta_1+\Delta_2)}+2e^{\beta(2\Delta_1+\Delta_2-2h)}+2e^{\beta(\Delta_1+2\Delta_2-2h)}+e^{\beta(3\Delta_2-4h)}$ \\
 \hline
 \end{tabular}
 \end{table}

 \begin{table}
 \caption{
    $Z_L$ for small values of $L$ using $a,b,c$ notation.  The case
   where the magnetic field is zero ($h=0$, $c=1$) is also shown.
     \label{tab:ZLabc} }
 \begin{tabular}{c c c}
 \hline
 $L$ & $Z_L(a,b,c)$ & $Z_L(a,b,c=1)$ \\
 \hline
 2 & $ac^2+2a+bc^{-2}$ & $3a+b$ \\
 3 & $a^2c^3+3a^2c+a^2c^{-1}+2abc^{-1}+b^2c^{-3}$ & $5a^2+2ab+b^2$ \\
 4 & $a^3c^4+4a^3c^2+3a^3+3a^2b+2a^2bc^{-2}+2ab^2c^{-2}+b^3c^{-4}$ & $8a^3+5a^2b+2ab^2+b^3$ \\
 \hline
 \end{tabular}
 \end{table}
 Though $\lambda_{\pm}$ contain a square root, it must be possible to express the $Z_L$ as finite
 polynomials in $a,b,c$.  We illustrate for small $L$ in tables
 \ref{tab:ZLDeltah} and \ref{tab:ZLabc}.
More generally,  $Z_L$ can be written as
\begin{linenomath*}
 \begin{align}
 Z_L& = \sum_{i=0}^{L-1} \sum_{j=0}^{L} \Omega_{L,i,j} e^{-\beta E_{L,i,j}}\\ 
 & = \sum_{i=0}^{L-1} \sum_{j=0}^{L} \Omega_{L,i,j} \exp [(L-1-i) \beta \Delta_1 + i \beta \Delta_2 + (L - 2j) \beta h ] \\
     & = \sum_{i=0}^{L-1} \sum_{j=0}^{L} \Omega_{L,i,j} a^{(L-1)-i}b^i c^{L-2j} \label{eq:ZLabc}. 
 \end{align}
\end{linenomath*}
where $\Omega_{L,i,j}$ is the number of states with energy $E_{L,i,j}=-\Delta_1(L-1-i)  - \Delta_2 i  - h (L-2j)$.
$\Omega_{L,i,j}$ could be calculated
by taking partial derivatives of the partition function in \ref{eq:ZLabcraw} with respect to $a$, $b$, and $c$ 
, setting those respective variables to zero and comparing with \ref{eq:ZLabc}
term by term giving
\begin{linenomath*}
 \begin{equation}
 \Omega_{L,i,j} = \left [ \left (\frac{1}{[(L-1)-i]!} \frac{\partial^{(L-1)-i}}{\partial a^{(L-1)-i}} \right )
    \left (\frac{1}{i!} \frac{\partial^i}{\partial b^i} \right ) \left (\frac{1}{(2j)!} \frac{\partial^{2j}}{\partial c^{2j}} \right )
     c^L Z_L(a,b,c) \right ]_{a=0,b=0,c=0}.
 \end{equation}
\end{linenomath*}
 However, this method is computationally expensive for large systems.  Instead we wrote the   partition function in \ref{eq:ZLabcraw} in the form 
\ref{eq:ZLabc} by algebraic rearrangement as described in detail in 
Appendix \ref{app:OmegaCalc} with the result:

\begin{linenomath*}
 \begin{equation}
 \Omega_{L,i,L-i-j}=\theta_{L,i,j}+\phi_{L,i,j}-\phi_{L,i-1,j},
 \end{equation}
\end{linenomath*}
 where
\begin{linenomath*}
\begin{gather}
\theta_{L,i,j} \equiv \sum_{k=j}^{\lfloor \frac{L+1}{2} \rfloor} \frac{2^{2j}}{2^{L+1}} \binom{L+1}{2k} \binom{k}{j}
             \sum_{l=0}^{2(k-j)} \binom{2(k-j)}{l} (-1)^l \binom{L+1-2k}{i-l}  \\
\phi_{L,i,j} \equiv \sum_{k=j}^{\lfloor \frac{L+1}{2} \rfloor} \frac{2^{2j}}{2^{L+1}} \binom{L+1}{2k+1} \binom{k}{j}
             \sum_{l=0}^{2(k-j)} \binom{2(k-j)}{l} (-1)^l \binom{L-2k}{i-l}
\end{gather}
\end{linenomath*}
When $h=0$, $c=1$ and  the result for the partition function simplifies to

\begin{linenomath*}
\begin{align}
 Z_L(h=0) & = \frac{\lambda_+^{L+1}}{a^2+(a-\lambda_+)^2} + \frac{\lambda_-^{L+1}}{a^2+(a-\lambda_-)^2} \label{eq:ZLh0gf} \\
     & = \sum_{i=0}^{L-1} \sum_{j=0}^{L}\Omega_{L,i,j} a^{(L-1)-i}b^i \label{eq:ZLab} \\
     & = \sum_{i=0}^{L-1} \Omega_{L,i} e^{-\beta E_{L,i}}.
\end{align}
\end{linenomath*}
Where $\Omega_{L,i}=\sum_j \Omega_{L,i,j}$ is the number of states with energy $E_{L,i}=-(L-1-i)\Delta_1- i\Delta_2 $,
\begin{linenomath*}
\begin{equation} \label{eq:h0eigval}
\lambda_\pm = \frac{1}{2} \left [ (a+b) \pm \sqrt{4a^2+(a-b)^2} \right ],
\end{equation}
\end{linenomath*}
and $a$ and $b$ retain the definitions  in equation \ref{eq:abcdef}.
$Z_L(h=0)$ is shown for small values of $L$  in table \ref{tab:ZLabc}.

$\Omega_{L,0}$ is the number of configurations when only $\Delta_1$ bonds are allowed.
A property of such configurations is that all sites with negative $\sigma$ have sites
with positive $\sigma$
as neighbors.  This property can be related to Fibonacci numbers
 \citep{honsberger1985mathematical}. The relation between $\Omega_{L,0}$ and the
Fibonacci numbers is described in detail in appendix \ref{app:NumTheor}.


For many Ising spin chains, 
we assume that the 
number of states associated with  a single chain of energy $E_{L,i,j}$ is 
$G_{L,i,j}=(V/v_L)\Omega_{L,i,j}$.  Here $\Omega_{L,i,j}$ is the number of states with energy $E_{L,i,j}$,
 $(V/v_L)$ is the number of places the chain
can be placed in volume $V$, and
$v_L$ is the volume occupied by  a polymer of length $L$. We take
$v_L$ to be related to the persistence length \citep{2018pub} ($l_p$) by 
$v_L=l_p^{3-3\nu}a^{3\nu} L^{3\nu}$  where $l_p$ is the polymer  persistence length \citep{2018pub} , $a$ is the bond length 
and $\nu$ is a dimensionless index . We write this as $v_L=v_pL^{3\nu}$ with $v_p =l_p^{3-3\nu}a^{3\nu}$. 
We used the value $\nu=1/2$ corresponding to random
walk behavior. (See also \citet{PRE2016} and \citet{2018pub}.)
A similar estimation for the number of states was used previously \citep{PRE2016}.
The partition function for chains of length $L$ of which there are $N_L=\sum_{i=0}^{L-1} \sum_{j=0}^L N_{L,i,j}$,
 where $N_{L,i,j}$ is the number of chains with energy $E_{L,i,j}$, can  be written as:
\begin{linenomath*}
\begin{equation} \label{eq:ZLNL}
Z_L(N_L)=\sum_{\sum_{i,j} N_{L,i,j}=N_L} \prod_{i=0}^{L-1} \ \prod_{j=0}^L \binom{N_{L,i,j}+G_{L,i,j}-1}{N_{L,i,j}} e^{-\beta E_{L,i,j} N_{L,i,j} },
\end{equation}
\end{linenomath*}
by an argument essentially identical to the one in reference \citet{PRE2016}.

For a system of many polymers of various lengths, the partition function
becomes
$Z(\{N_L\})=\prod_{L=1}^{L_{\text{max}}} Z_L(N_L)$ or using the previous expression:
\begin{linenomath*}
\begin{equation} \label{eq:ZNL}
Z(\{N_L\})=\prod_{L=1}^{L_{\text{max}}} \sum_{\{N_{L,i,j}\} \ni\sum_{i,j} N_{L,i,j}=N_L} \prod_{i=0}^{L-1} \ \prod_{j=0}^L \binom{N_{L,i,j}+G_{L,i,j}-1}{N_{L,i,j}} e^{-\beta E_{L,i,j} N_{L,i,j} }.
\end{equation}
\end{linenomath*}
Denoting the total number of chains as $N=\sum_{L=1}^{L_{\text{max}}} N_L$ we can then write the
grand canonical partition function $\mathcal{Z}$ as:
\begin{linenomath*}
\begin{equation} \label{eq:GrandZ}
\mathcal{Z}=\sum_{N=0}^\infty \sum_{\{N_L\} \ni \sum_LN_L=N} Z(\{N_L\})=
\sum_{\{N_L\}}Z(\{N_L\}) e^{\mu \sum_L\beta N_L}
\end{equation}
\end{linenomath*}
where the sum on $\{N_L\} $ in the last expression is unrestricted.
Expanding $Z(\{N_L\})$, then $Z_L(N_L)$, and using the definition
that $N=\sum_{L=1}^{L_{\text{max}}} N_L =\sum_{L=1}^{L_{\text{max}}} \sum_{i=0}^{L-1} \sum_{j=0}^L N_{L,i,j}$,
we can move the summation over $\{N_L\}$ into the products with respect to $L$, $i$, and $j$.
and also remove the restriction on the  sum
 $\sum_{\{N_{L,i,j}\} \ni \sum_{i,j} N_{L,i,j}=N_L}$ yielding:
\begin{linenomath*}
\begin{align} \label{eq:grandcon}
\mathcal{Z} & = \prod_{L=1}^{L_{\text{max}}} \ \prod_{i=0}^{L-1} \ \prod_{j=0}^L \ \sum_{N_{L,i,j}=0}^\infty \ \binom{N_{L,i,j}+G_{L,i,j}-1}{N_{L,i,j}} e^{(\mu \beta -\beta E_{L,i,j}) N_{L,i,j} } \\
            & = \prod_{L=1}^{L_{\text{max}}} \ \prod_{i=0}^{L-1} \ \prod_{j=0}^L \ \left ( \frac{1}{1-\exp (\mu \beta -\beta E_{L,i,j})} \right )^{G_{L,i,j}}
\end{align}
\end{linenomath*}

where in the last equality we used the identity
\begin{linenomath*}
\begin{equation}
\sum_{n=k}^\infty \binom{n}{k} y^n = \frac{y^k}{(1-y)^{k+1}}
\end{equation}
\end{linenomath*}
with $k=G_{L,i,j}+1, n=N_{L_{ij}}+G_{L_{ij}}-1, y=e^{(\mu-E_{L,i,j})\beta}$

The Helmholtz free energy is then proportional to 
\begin{linenomath*}
\begin{equation} \label{eq:lnBigZ}
\ln \mathcal{Z} = \sum_{L=1}^{L_{\text{max}}} \ \sum_{i=0}^{L-1} \ \sum_{j=0}^L \ G_{L,i,j}
          \ln \left ( \frac{1}{1-\exp (\tilde{\mu} -\beta E_{L,i}+(L-2j)\tilde{h})} \right )
\end{equation}
\end{linenomath*}
where we denote $\tilde{\mu} \equiv \beta \mu$, $\tilde{h} \equiv \beta h$,
 and $E_{L,i} \equiv -\Delta_1(L-1-i)- \Delta_2 i$ 
The following quantities can then be calculated by taking partial derivatives 
$\ln \mathcal{Z}$
of equation \ref{eq:lnBigZ}:
\begin{linenomath*}
\begin{align}
\text{Expected Total Number of Chains:} \ & \ \langle N \rangle = 
\left (\frac{\partial}{\partial \tilde{\mu}} \ln \mathcal{Z} \right )_{\beta,\tilde{h}}\label{eq:Np} \\
\text{Expected Total Energy:}             \ & \ \langle E \rangle = - \left (\frac{\partial}{\partial \beta} \ln \mathcal{Z}\right )_{\tilde{\mu},\tilde{h}} \label{eq:Ener} \\
\text{Expected Monomer Type Imbalance:} \ & \ \langle N_+-N_- \rangle = \left (\frac{\partial}{\partial \tilde{h}} \ln \mathcal{Z} \right )_{\beta,\tilde{\mu}}\label{eq:NcmNn}
\end{align}
\end{linenomath*}
where $N_\pm$ is the total number of sites with $\sigma =\pm 1$ respectively.
Note that when calculating the energy $E$,  $\tilde{\mu}$ and $\tilde{h}$
are fixed (ie they are not regarded as $\beta $ dependent.)
That is because the energy of interest is  only  the energy associated with bonds and not
the energy associated with the chemical potential and the artificial magnetic field. 


In equation \ref{eq:lnBigZ} if the value of
$(\tilde{\mu} -\beta E_{L,i}+(L-2j)\tilde{h})$
is large and negative the following approximation, which we call the Gibbs limit,
can be used:
\begin{linenomath*}
\begin{equation}
\ln \left ( \frac{1}{1-\exp (\tilde{\mu} -\beta E_{L,i}+(L-2j)\tilde{h})} \right )
  \approx \exp (\tilde{\mu} -\beta E_{L,i}+(L-2j)\tilde{h}).
\end{equation}
\end{linenomath*}

Then $\ln \mathcal{Z}$ becomes:
\begin{linenomath*}
\begin{align}
\ln \mathcal{Z} & \approx  \sum_{L=1}^{L_{\text{max}}} \ \sum_{i=0}^{L-1} \ \sum_{j=0}^L \ G_{L,i,j} 
                              \exp (\tilde{\mu} -\beta E_{L,i}+(L-2j)\tilde{h}) \\
                & =  e^{\tilde{\mu}} \sum_{L=1}^{L_{\text{max}}} \
                     (V/v_L) \left ( \sum_{i=0}^{L-1} \ \sum_{j=0}^L \Omega_{L,i,j} \exp (\tilde{\mu} -\beta E_{L,i}+(L-2j)\tilde{h}) \right ) \\
                & =  e^{\tilde{\mu}} \sum_{L=1}^{L_{\text{max}}} (V/v_L) Z_L (\tilde{h},\beta) \label{eq:GibbsApprox}
\end{align}
\end{linenomath*}
Using this approximate form we have:
\begin{linenomath*}
\begin{equation} 
\langle N \rangle = \left \langle \sum_{L=1}^{L_{\text{max}}} N_L \right \rangle = \sum_{L=1}^{L_{\text{max}}} \langle N_L \rangle
          \approx \frac{\partial}{\partial \tilde{\mu}} e^{\tilde{\mu}} \sum_{L=1}^{L_{\text{max}}} (V/v_L) Z_L
               = \sum_{L=1}^{L_{\text{max}}} \ e^{\tilde{\mu}} (V/v_L) Z_L
\end{equation}
\end{linenomath*}
so that 
\begin{linenomath*}
\begin{align}
\text{Expected Number of Chains of length $L$=} \ & \ \langle N_L \rangle = e^{\tilde{\mu}}  (V/v_L) Z_L = e^{\tilde{\mu}}  \frac{V}{v_p L^{3/2}} Z_L \label{eq:NL} \\
\text{Expected Total Number of Monomers:}          \ & \ \langle N_++N_- \rangle = e^{\tilde{\mu}} \sum_{L=1}^{L_{\text{max}}} \ L (V/v_L) Z_L \label{eq:Nm} \\
\text{Expected Monomer Type Imbalance $m=$:}                  \ & \ \langle m \rangle = \frac{ \langle N_+-N_- \rangle }{ \langle N_++N_- \rangle }
                                             = \frac{ \sum_{L=1}^{L_{\text{max}}} \frac{\partial}{\partial \tilde{h}} Z_L / v_L }{\sum_{L=1}^{L_{\text{max}}} \ L Z_L / v_L}
                                                \label{eq:M}
\end{align}
\end{linenomath*}

\section{Application to Titan Data} \label{sec:Results}

Atmospheric data from Titan
\citep{desai2017carbon} was analyzed to 
extract a length distribution as described in \citet{2018pub}, 
assuming that all the detected molecules were linear polymers.
To compare the length distributions inferred from the data with
the ones expected in equilibrium we established that the Gibbs limit 
was a good approximation and used \ref{eq:NL} rearranged as
\begin{linenomath*}
\begin{equation}
\frac{\langle N_L \rangle v_p}{V} = \langle \rho_L \rangle v_p = e^{\tilde{\mu}} \ Z_L / L^{3/2}
\end{equation}
\end{linenomath*}
where $v_p=l_p^{3-3\nu}a^{3\nu}$   and $\rho_L=N_L/V$  
is the volume density of chains of length $L$ so that  the total 
density is  $\rho = \sum_{L=1}^{L_{\text{max}}} \rho_L$. 
An energy density $u = \langle E \rangle /V$ can also be extracted
as described in  \citet{2018pub}. 
We then proceed as follows:
Set the experimentally determined number, energy densities and monomer type 
imbalance $m$ to the equilibrium expressions and solve the resulting implicit 
equations for $\tilde{\mu}, \tilde{h} $ and $\beta$ numerically with solutions
denoted $\tilde{\mu}(\rho,u,m),\tilde{h}(\rho,u,m),\beta(\rho,u,m)$.  (We
used $m=-0.98$ assuming that 2\% of the monomers in  the chains are 
carbon (positive $\sigma$.) )Using those
values in the expression \ref{eq:NL} gives what we call the local
equilibrium value $\overline{N_L(\tilde{\mu}(\rho,u,m),\tilde{h}(\rho,u,m),\beta(\rho,u,m))}$
for each $L$.  The experimental values of the $N_L$ differ from these values
because the Titan atmosphere is not in local equilibrium.  We measure the 
degree to which it is out of local equilibrium by a normalized  Euclidean distance 
$R_L$ between
the local equilibrium point and the experimental point in  the space $\{N_L\}$ of 
populations of polymers of various lengths $L$.  The space has a 
dimension of up to 10$^4$, though only values up to about  $L \approx 10^3$ are 
numerically significant.  Specifically
%
\begin{linenomath*}
\begin{equation}
R_L = \sqrt{
  \sum_L (v_p/V)^2 (N_L-\overline{N_L(\tilde{\mu}(\rho,u,m),\tilde{h}(\rho,u,m),\beta(\rho,u,m))})^2
}/(v_p \rho \sqrt{2}),
\end{equation}
\end{linenomath*}
where $N_L$ is the length distribution of the data set.

We made a similar determination of an equilibrium point in the space $\{N_L\}$ corresponding
to equilibrium with an external heat bath with a fixed $\beta$ value.  In that case, 
we set the experimental values of the number density $\rho$ and the monomer imbalance 
$m$ to their equilibrium expressions and solved the resulting implicit equations
for $\tilde{\mu} $ and $\tilde{h}$ numerically while leaving $\beta$ fixed. 
(The value 
T= $120$ degrees Kelvin \citep{crary2009heavy} was used to fix $\beta=1/k_BT$.)
The resulting values of $\tilde{\mu}(\rho,m),\tilde h(\rho,m)$  were then inserted 
in the equilibrium expressions giving the coordinates of a point in  the 
space $\{N_L\}$ described by
$\overline{N_L(\tilde{\mu}(\rho,\beta,m),\tilde{h}(\rho,\beta,m))}$
We then evaluate a second normalized Euclidean distance from that equilibrium point,
termed the 'thermal' equilibrium point as
\begin{linenomath*}
\begin{equation}
R_T = \sqrt{
  \sum_L (v_p/V)^2 (N_L-\overline{N_L(\tilde{\mu}(\rho,\beta,m),\tilde{h}(\rho,\beta,m))})^2
}/(v_p \rho \sqrt{2}),
\end{equation}
\end{linenomath*}
where $N_L$ is the length distribution of the experimental data set.

\begin{figure}
 \includegraphics{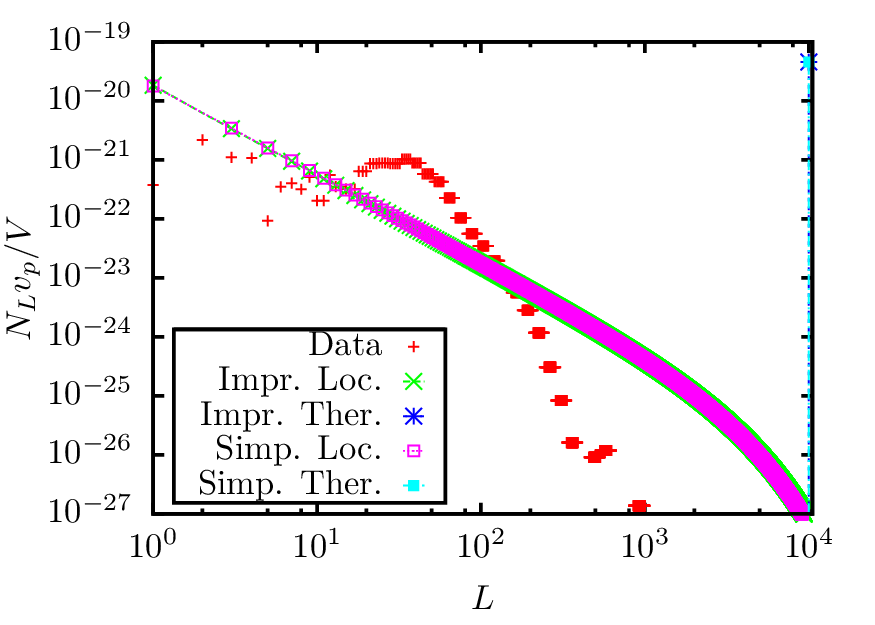}%
 \caption{$N_L v_p/V$ values for the Titan atmosphere at an altitude of 1013km.
  The data as well as the calculated improved (Impr.) and simple (Simp.)
    local (Loc.) and thermal (Therm.) equilibria are shown.
   \label{fig:titan1013}}%
 \end{figure}

\begin{figure}
 \includegraphics{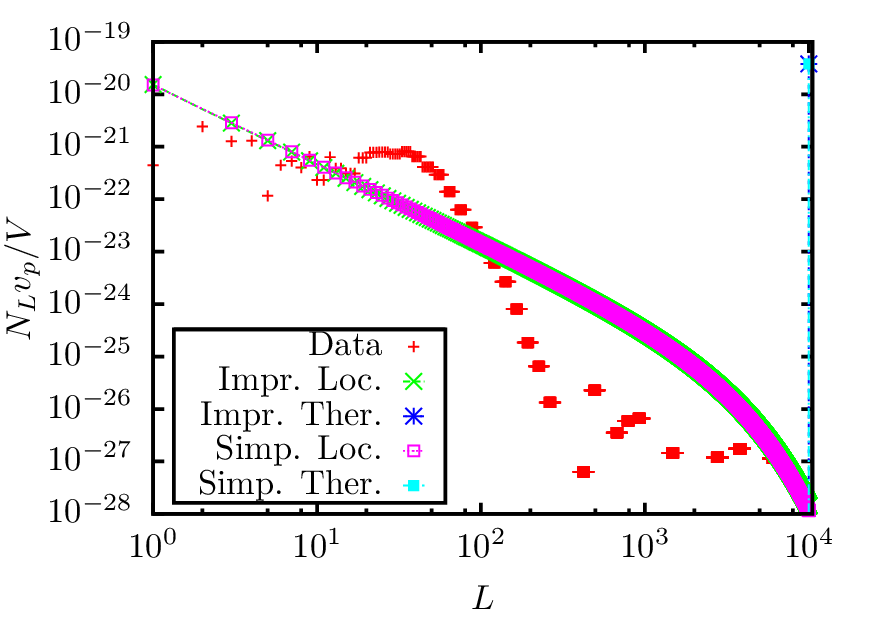}%
 \caption{$N_L v_p/V$ values for the Titan atmosphere at an altitude of 1032km.
  The data as well as the calculated improved (Impr.) and simple (Simp.)
    local (Loc.) and thermal (Therm.) equilibria are shown.
   \label{fig:titan1032}}%
 \end{figure}

\begin{figure}
 \includegraphics{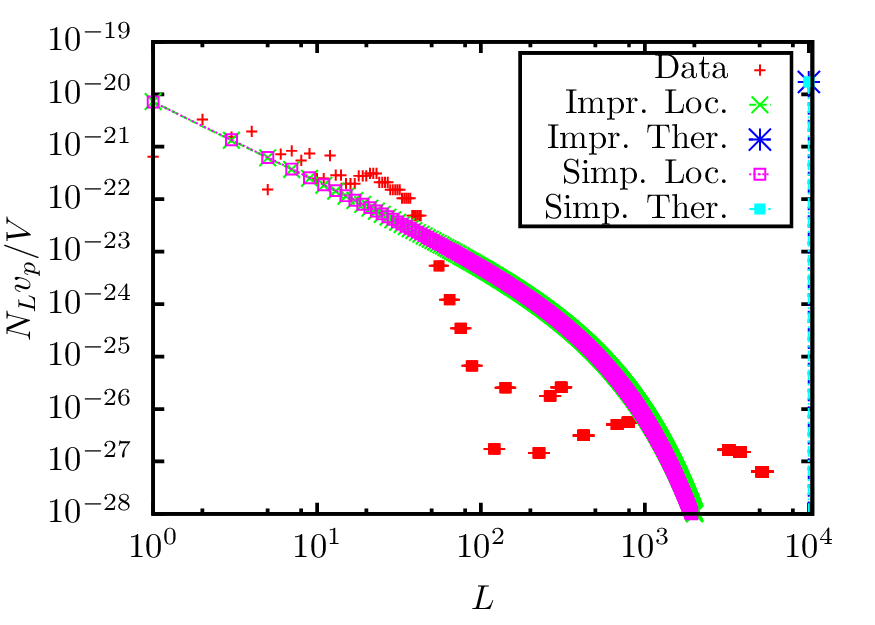}%
 \caption{$N_L v_p/V$ values for the Titan atmosphere at an altitude of 1078km.
  The data as well as the calculated improved (Impr.) and simple (Simp.)
    local (Loc.) and thermal (Therm.) equilibria are shown.
   \label{fig:titan1078}}%
 \end{figure}

\begin{figure}
 \includegraphics{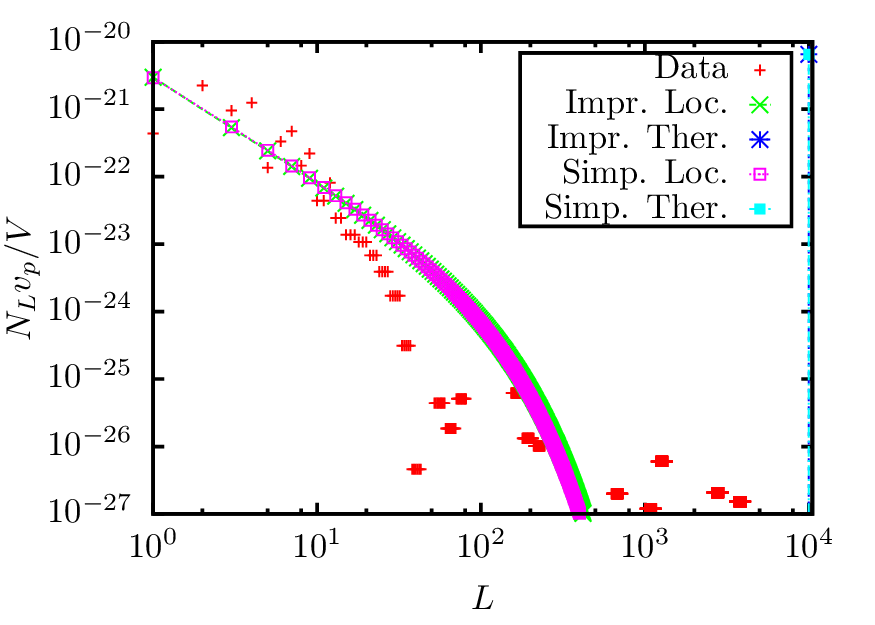}%
 \caption{$N_L v_p/V$ values for the Titan atmosphere at an altitude of 1148km.
  The data as well as the calculated improved (Impr.) and simple (Simp.)
    local (Loc.) and thermal (Therm.) equilibria are shown.
   \label{fig:titan1148}}%
 \end{figure}

\begin{figure}
 \includegraphics{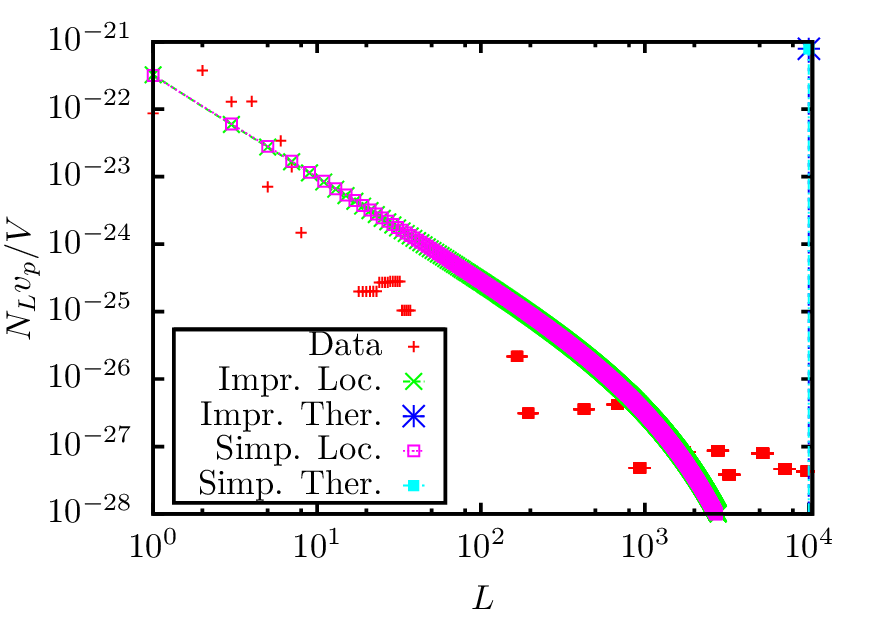}%
 \caption{$N_L v_p/V$ values for the Titan atmosphere at an altitude of 1244km.
  The data as well as the calculated improved (Impr.) and simple (Simp.)
    local (Loc.) and thermal (Therm.) equilibria are shown.
   \label{fig:titan1244}}%
 \end{figure}

Figures \ref{fig:titan1013}-\ref{fig:titan1244}
show the distributions of $N_L v_p/V$ for  various
altitude measurements and table \ref{tab:TitanValues}
shows the numerical results  for the parameters characterizing the local and
thermal equilibrium points.


 \begin{table}
 \caption{
   Values from the equilibrium calculations performed using  the Titan
   data and the resulting values of $R_L$ and $R_T$.   \label{tab:TitanValues} }
 \begin{tabular}{ c c c c c c c c c}
 \hline
 Altitude (km) & $R_L$    & Local $\tilde{\mu}$ & Local $\Delta_1 \beta$ & Local $\tilde{h}$ & $R_T$   & Thermal $\tilde{\mu}$ & Thermal $\Delta_1 \beta$ & Thermal $\tilde{h}$ \\
 \hline
 $1013$        & $0.299$  & $-46.2$             & $-2.05$                & $-0.992$          & $0.713$ & $-3.28 \times 10^6$            & $326$                    & $-167$              \\
 $1032$        & $0.293$  & $-46.3$             & $-2.06$                & $-0.997$          & $0.714$ & $-3.28\times 10^6$            & $326$                    & $-167$              \\
 $1078$        & $0.284$  & $-47.1$             & $-2.17$                & $-1.05$           & $0.732$ & $-3.28\times 10^6$            & $326$                    & $-167$              \\
 $1148$        & $0.328$  & $-48.1$             & $-2.36$                & $-1.14$           & $0.771$ & $-3.28\times 10^6$            & $326$                    & $-167$              \\
 $1244$        & $0.339$  & $-50.2$             & $-2.10$                & $-1.01$           & $0.805$ & $-3.28\times 10^6$            & $326$                    & $-167$              \\
 \hline
 \end{tabular}
 \end{table}

\section{Discussion and Conclusions} \label{sec:DiscAndConc}

As one can see from the figures, the results of the improved 
calculation of the partition function presented here are in excellent
agreement with the results of the simple mean field approximation 
used in \citet{2018pub}.  This appears to be mainly because the atomic 
fraction of carbon in the application is very small (2\%), making
corrections to a model with uniform bond strength small. It appears,
however that the lowest order corrections to the $p \rightarrow 0$
limit in the two solutions are not the same.  It would be interesting
to explore this aspect of the two approaches further.


%
%
%
%

\appendix

\section{Calculations of $\Omega_{L,i,j}$.} \label{app:OmegaCalc}

Here we describe the algebraic rearrangement of   equation \ref{eq:ZLabcraw} which gives  the  form  closed form \ref{eq:ZLabc} 
for  $\Omega_{L,i,j}$. 
We consider a slightly different model in which  
the magnetic field term
 is defined as:
\begin{linenomath*}
 \begin{equation}
 -h' \sum_{i=1}^L \frac{1}{2}(\sigma_i+1)
 \end{equation}
\end{linenomath*}
 and then relate the coefficients of an expansion of the partition function in that
 model to the coefficients in the original model.
 Notice that $h'$  counts the number of sites with $\sigma_i=+1$.  
 Going through the same calculations described  in section II 
 yields the partition function and eigenvalues
\begin{linenomath*}
 \begin{gather}
  Z_L'(a,b,c') = \frac{\lambda_+^{L+1}}{a^2c'+(ac'-\lambda_+)^2} + \frac{\lambda_-^{L+1}}{a^2c'+(ac'-\lambda_-)^2}
       =\sum_{i=0}^{L-1} \sum_{j=0}^L \Omega_{L,i,j} a^{(L-1)-i} b^i c'^j \label{eq:ZLhcraw} \\
  \lambda_\pm = \frac{1}{2} \left [ (ac'+b) \pm \sqrt{4a^2c'+(ac'-b)^2} \right ]
 \end{gather}
\end{linenomath*}
 where $a=exp(\beta \Delta_1 )$, $b= exp(\beta \Delta_2)$ (as in the main text) and $c'=exp(\beta h')$
The eigenvalues and the partition function are different in the factors involving
the field $h'$ because of the different field term.
Note that in  \ref{eq:ZLhcraw} the RHS contains no radicals, whereas the middle equation contains radicals.
  Secondly the RHS contains no denominator, so at some point the denominator is factored out
from the numerator.
 In the following  the binomial theorem is used frequently:
\begin{linenomath*}
\begin{equation}
(x+y)^n=\sum_{k=0}^{n} \binom{n}{k} x^{n-k}y^{k} \ , \ \binom{n}{k}=\frac{n!}{k!(n-k)!}
\end{equation}
\end{linenomath*}
 $Z_L'$ is rearranged as 
\begin{linenomath*}
\begin{align}
Z_L'(a,b,c') &= \frac{\lambda_+^{L+1}}{a^2c'+(ac'-\lambda_+)^2}+\frac{\lambda_-^{L+1}}{a^2c'+(ac'-\lambda_-)^2} \\
 & = \frac{\lambda_+^{L+1}[a^2c'+(ac'-\lambda_-)^2]+\lambda_-^{L+1}[a^2c'+(ac'-\lambda_+)^2]}{[a^2c'+(ac'-\lambda_+)^2][a^2c'+(ac'-\lambda_-)^2]} \label{eq:za1numden}
\end{align}
\end{linenomath*}
and the eigenvalues as  $\lambda_\pm$:
\begin{linenomath*}
\begin{gather}
  \lambda_\pm = \frac{1}{2} (\eta \pm \delta) \\
  \eta \equiv (ac'+b) \\
  \delta \equiv \sqrt{4a^2c'+(ac'-b)^2}
\end{gather}
\end{linenomath*}
We then simplify $\lambda_\pm^{L+1}$ by separating  its non-radical and radical terms:
\begin{linenomath*}
\begin{align}
 \lambda_\pm^{L+1} & = \left [ \frac{1}{2} (\eta \pm \delta) \right ]^{L+1} \\
 & = \frac{1}{2^{L+1}} (\eta \pm \delta)^{L+1} \\
 & = \frac{1}{2^{L+1}} \sum_{i=0}^{L+1} \binom{L+1}{i} \eta^{(L+1)-i} (\pm \delta)^i \\
 & = \frac{1}{2^{L+1}} \sum_{i \ \text{even}} \binom{L+1}{i} \eta^{(L+1)-i} \delta^i
      \pm \delta \frac{1}{2^{L+1}} \sum_{i \ \text{odd}} \binom{L+1}{i} \eta^{(L+1)-i} \delta^{i-1} \\
 & = \frac{1}{2^{L+1}} \sum_{i=0}^{\lfloor \frac{L+1}{2} \rfloor} \binom{L+1}{2i} \eta^{(L+1)-2i} \delta^{2i}
       \pm \delta \frac{1}{2^{L+1}} \sum_{i=0}^{\lfloor \frac{L+1}{2} \rfloor} \binom{L+1}{2i+1} \eta^{L-2i} \delta^{2i} \\
 & = x \pm \delta y
\end{align} 
\end{linenomath*}
 where:
\begin{linenomath*}
\begin{align}
x & \equiv \frac{1}{2^{L+1}} \sum_{i=0}^{\lfloor \frac{L+1}{2} \rfloor} \binom{L+1}{2i} \eta^{(L+1)-2i} \delta^{2i} \\
  & = \frac{1}{2^{L+1}} \sum_{i=0}^{\lfloor \frac{L+1}{2} \rfloor} \binom{L+1}{2i} [4a^2c'+(ac'-b)^2]^i(ac'+b)^{(L+1)-2i} \\
y & \equiv \frac{1}{2^{L+1}} \sum_{i=0}^{\lfloor \frac{L+1}{2} \rfloor} \binom{L+1}{2i+1} \eta^{L-2i} \delta^{2i} \\
  & = \frac{1}{2^{L+1}} \sum_{i=0}^{\lfloor \frac{L+1}{2} \rfloor} \binom{L+1}{2i+1} [4a^2c'+(ac'-b)^2]^i(ac'+b)^{L-2i}
\end{align}
\end{linenomath*}
We also simplify the terms in the square brackets in equation \ref{eq:za1numden}:
\begin{linenomath*}
\begin{gather}
  [a^2c'+(ac'-\lambda_+)^2]=2a^2c'+\frac{1}{2}(ac'-b)[(ac'-b)-\delta] \\
  [a^2c'+(ac'-\lambda_-)^2]=2a^2c'+\frac{1}{2}(ac'-b)[(ac'-b)+\delta] \\
  [a^2c'+(ac'-\lambda_-)^2][a^2c'+(ac'-\lambda_+)^2]=a^2c'\delta^2
\end{gather}
\end{linenomath*}
in this notation  $Z_L'$ becomes:
\begin{linenomath*}
\begin{align}
Z_L'(a,b,c') & = \frac{\lambda_+^{L+1}[a^2c'+(ac'-\lambda_-)^2]+\lambda_-'^{L+1}[a^2c'+(ac'-\lambda_+)^2]}{[a^2c'+(ac'-\lambda_+)^2][a^2c'+(ac'-\lambda_-)^2]} \\
 & = \frac{
      (x+\delta y)(2a^2c'+\frac{1}{2}(ac'-b)[(ac'-b)+\delta])+(x-\delta y)(2a^2c'+\frac{1}{2}(ac'-b)[(ac'-b)-\delta])
      }{a^2c'\delta^2} \\
 & = \frac{[4a^2c'+(ac'-b)^2]x+(ac'-b)y\delta^2}{a^2c'\delta^2} \\
 & = \frac{\delta^2x+(ac'-b)y\delta^2}{a^2c'\delta^2} \\
 & = \frac{x+(ac'-b)y}{a^2c'} \label{eq:zlxy}
\end{align}
\end{linenomath*}
 $x$ and $y$ are rewritten  to give a series in powers of  $a$, $b$, and $c'$:
\begin{linenomath*}
\begin{align}
x & =  \frac{1}{2^{L+1}} \sum_{i=0}^{\lfloor \frac{L+1}{2} \rfloor} \binom{L+1}{2i} [4a^2+(ac'-b)^2]^i(ac'+b)^{(L+1)-2i} \\
  & =  \frac{1}{2^{L+1}} \sum_{i=0}^{\lfloor \frac{L+1}{2} \rfloor} \binom{L+1}{2i} \nonumber \\
  & \ \  \times  \left [ \sum_{j=0}^i \binom{i}{j} (4a^2c')^j \sum_{k=0}^{2(i-j)} \binom{2(i-j)}{k}(-ac')^{2(i-j)-k} b^k \right ] \nonumber \\
  & \ \  \times  \left [  \sum_{l=0}^{L+1-2i} \binom{L+1-2i}{l}(ac)^{L+1-2i} b^l \right ] \\
  & = \sum_{i,j,k,l}
      \frac{2^{2j}}{2^{L+1}}
      \binom{L+1}{2i} \binom{i}{j} \binom{2(i-j)}{k}(-1)^k \binom{L+1-2i}{l} a^{(L+1)-(l+k)} b^{l+k} c'^{(L+1)-(l+k)-j}
\end{align}
\end{linenomath*}
We rewrite this in terms of the summation variable   $m=l+k$ instead of $l$.  The limits on the 
summation on $m$ are somewhat complicated but we show that the the substitution  
 $\sum_{l=0}^{L+1-2i} \rightarrow \sum_{m=0}^{L+1}$ is justified because the extension of the limits
 on $m$ only adds terms which are zero.  The order of the sums on $i$ and $j$ is also
 swapped $\sum_{i=0}^{\lfloor \frac{L+1}{2} \rfloor} \sum_{j=0}^i \rightarrow \sum_{j=0}^{\lfloor \frac{L+1}{2} \rfloor} \sum_{i=j}^{\lfloor \frac{L+1}{2} \rfloor}$
 yielding the following form for  $x$ 
\begin{linenomath*}
 \begin{equation}
 x=\sum_{m=0}^{L+1} \sum_{j=0}^{\lfloor \frac{L+1}{2} \rfloor} \theta_{L,m,j} a^{(L+1)-m} b^m c'^{(L+1)-m-j}
 \end{equation}
\end{linenomath*}
where
\begin{linenomath*}
\begin{equation}
\theta_{L,m,j} \equiv \sum_{i=j}^{\lfloor \frac{L+1}{2} \rfloor} \frac{2^{2j}}{2^{L+1}} \binom{L+1}{2i} \binom{i}{j} \sum_{k=0}^{2(i-j)} \binom{2(i-j)}{k}(-1)^k \binom{L+1-2i}{m-k}
\end{equation}
\end{linenomath*}
Similarly $y$ can be written as:
\begin{linenomath*}
 \begin{equation}
 y=\sum_{m=0}^{L} \sum_{j=0}^{\lfloor \frac{L+1}{2} \rfloor} \phi_{L,m,j} a^{L-m} b^m c'^{L-m-j}
 \end{equation}
\end{linenomath*}
where:
\begin{linenomath*}
\begin{equation}
\phi_{L,m,j} \equiv \sum_{i=j}^{\lfloor \frac{L+1}{2} \rfloor} \frac{2^{2j}}{2^{L+1}} \binom{L+1}{2i+1} \binom{i}{j} \sum_{k=0}^{2(i-j)} \binom{2(i-j)}{k}(-1)^k \binom{L-2i}{m-k}
\end{equation}
\end{linenomath*}
These expressions for $x$ and $y$ are then inserted into equation \ref{eq:zlxy} and the sums are rearranged
\begin{linenomath*}
\begin{align}
Z_L'(a,b,c') & = \frac{x+(ac'-b)y}{a^2c'} \\
 & = \frac{1}{a^2c'} \sum_j \left [ \sum_m \theta_{L,m,j} a^{(L+1)-m} b^m c'^{(L+1)-m-j} + (ac'-b) \sum_m \phi_{L,m,j} a^{L-m} b^m c'^{L-m-j} \right ] \\
 & = \sum_j \left [ \sum_m \theta_{L,m,j} a^{(L-1)-m} b^m c'^{L-m-j} + (\frac{1}{a}-\frac{b}{a^2c'}) \sum_m \phi_{L,m,j} a^{L-m} b^m c'^{L-m-j} \right ] \\
 & = \sum_j \Bigg [ \sum_m \theta_{L,m,j} a^{(L-1)-m} b^m c'^{L-m-j} + \sum_m \phi_{L,m,j} a^{(L-1)-m} b^m c'^{L-m-j} \nonumber  \\
 &  \ \ -  \sum_m \phi_{L,m,j} a^{(L-2)-m} b^{m+1} c'^{L-1-m-j}  \Bigg ] \\
 & = \sum_j \left [ \sum_m (\theta_{L,m,j}+\phi_{L,m,j}) a^{(L-1)-m} b^m c'^{L-m-j} - \sum_m \phi_{L,m-1,j} a^{(L-1)-m} b^m c'^{L-m-j} \right ] \\
 & = \sum_{m=0}^{L+1} \sum_{j=0}^{\lfloor \frac{L+1}{2} \rfloor} (\theta_{L,m,j}+\phi_{L,m,j}-\phi_{L,m-1,j}) a^{(L-1)-m} b^m c'^{L-m-j} \\
 & = \sum_{i=0}^{L-1} \sum_{j=0}^L \Omega_{L,i,j} \ a^{(L-1)-i} b^i c'^j \label{eq:zluv}
\end{align}
\end{linenomath*}
In going from the form (A38) to (A39) we introduced a change of summation 
variable $m'=m-1$ which changes the lower limit from $m=0$ to $m'=1$.
However the term with $m'=0$ is zero and can be formally included.
By comparing powers of $a$, $b$, and $c'$ we then have:
\begin{linenomath*}
\begin{equation} \label{eq:ouv}
\Omega_{L,m,L-m-j}' = \theta_{L,m,j}+\phi_{L,m,j}-\phi_{L,m-1,j}
\end{equation}
\end{linenomath*}
where it has been numerically verified that the coefficients of the
 $b^L$ and $b^{L+1}$ terms are zero.

 Finally we relate $\Omega_{L,m,L-m-j}'$ to the corresponding quantity $\Omega_{L,m,L-m-j}$ in the original model
 of the main text by relating the partition functions:

Let
\begin{linenomath*}
\begin{gather}
Z_L(h,\beta)=\sum_{\boldsymbol{\sigma}} \exp [-\beta H(\boldsymbol{\sigma})] \\
H(\boldsymbol{\sigma})=-\sum_{i=1}^{L-1} J(\sigma_i,\sigma_{i+1}) \, \sigma_i \sigma_{i+1}-h \sum_{i=1}^{L} \sigma_i
\end{gather}
\end{linenomath*}
which is the canonical form of the partition function of a spin system with
an external magnetic field $h$ as described in  section II.  Note that value of the term $\sum_i \sigma_i$ is the spin difference
$N_+-N_-$ which ranges from $L$ to $-L$ in steps of 2 ($-L,-L+2,-L+4,\cdots,L-4,L-2,L$).

Let 
\begin{linenomath*}
\begin{gather}
Z'_L(h',\beta)=\sum_{\boldsymbol{\sigma}} \exp [-\beta H'(\boldsymbol{\sigma})] \\
H'(\boldsymbol{\sigma})=-\sum_{i=1}^{L-1} J(\sigma_i,\sigma_{i+1}) \, \sigma_i \sigma_{i+1}-\frac{h'}{2} \sum_{i=1}^{L} (\sigma_i+1)
\end{gather}
\end{linenomath*}
where now the term $\frac{1}{2} \sum_{i=1}^{L} (\sigma_i+1)$ counts the number of positive spins $N_+$ which ranges from $0$ to $L$.

To find a relation between $Z_L$ and $Z'_L$ we rearrange  $Z'_L$:
\begin{linenomath*}
\begin{align}
Z'_L(h',\beta) & = \sum_{\boldsymbol{\sigma}} \exp \left [ \beta \sum_{i=1}^{L-1} J(\sigma_i,\sigma_{i+1}) \, \sigma_i \sigma_{i+1}+\frac{\beta h'}{2} \sum_{i=1}^{L} (\sigma_i+1) \right ] \\
 & = \exp \left [ \frac{L \beta h'}{2} \right ] \sum_{\boldsymbol{\sigma}} \exp \left [ \beta \sum_{i=1}^{L-1} J(\sigma_i,\sigma_{i+1}) \, \sigma_i \sigma_{i+1}+\frac{\beta h'}{2} \sum_{i=1}^{L} \sigma_i \right ] \\
\end{align}
\end{linenomath*}
Now let $h'=2h$:
\begin{linenomath*}
\begin{align}
Z'_L(2h,\beta)  & = \exp \left [ L \beta h \right ] \sum_{\boldsymbol{\sigma}} \exp \left [ \beta \sum_{i=1}^{L-1} J(\sigma_i,\sigma_{i+1}) \, \sigma_i \sigma_{i+1}+\beta h \sum_{i=1}^{L} \sigma_i \right ] \\
 & = \exp \left [ L \beta h \right ] \sum_{\boldsymbol{\sigma}} \exp [-\beta H(\boldsymbol{\sigma})] \\
 & = \exp \left [ L \beta h \right ] Z_L(h,\beta)
\end{align}
\end{linenomath*}

$Z_L$ and $Z'_L$ can be written in the form:
\begin{linenomath*}
\begin{gather}
Z_L= \sum_{i=0}^{L-1} \sum_{j=0}^{L} \Omega_{L,i,j} a^{(L-1)-i}b^i c^{L-2j} \\
Z'_L= \sum_{i=0}^{L-1} \sum_{j=0}^L \Omega'_{L,i,j} \ a^{(L-1)-i} b^i c'^j
\end{gather}
\end{linenomath*}
where
\begin{linenomath*}
\begin{equation}
a \equiv \exp (\beta \Delta_1) \ , \ b \equiv \exp (\beta \Delta_2) \ , \ c \equiv \exp (\beta h) \ , \ c' \equiv \exp(\beta h').
\end{equation}
\end{linenomath*}

By using the relation $Z'_L(h'=2h,\beta)=\exp \left [ L \beta h \right ] Z_L(h,\beta)$, (thus $c'=c^2$) we  relate $\Omega$ and $\Omega'$:
\begin{linenomath*}
\begin{gather}
Z'_L(h'=2h,\beta)=\exp \left [ L \beta h \right ] Z_L(h,\beta) \\
\sum_{i=0}^{L-1} \sum_{j=0}^L \Omega'_{L,i,j} \ a^{(L-1)-i} b^i (c^2)^j = c^L \sum_{i=0}^{L-1} \sum_{j=0}^{L} \Omega_{L,i,j} a^{(L-1)-i}b^i c^{L-2j} \\
\sum_{i=0}^{L-1} \sum_{j=0}^L \Omega'_{L,i,j} \ a^{(L-1)-i} b^i c^{2j} = \sum_{i=0}^{L-1} \sum_{j=0}^{L} \Omega_{L,i,j} a^{(L-1)-i}b^i c^{2(L-j)} \\
\sum_{i=0}^{L-1} \sum_{j=0}^L \Omega'_{L,i,j} \ a^{(L-1)-i} b^i c^{2j} = \sum_{i=0}^{L-1} \sum_{j=0}^{L} \Omega_{L,i,L-j} a^{(L-1)-i}b^i c^{2j}
\end{gather}
\end{linenomath*}
Therefore $\Omega'_{L,i,j} = \Omega_{L,i,L-j}$.


\section{Relation to Fibonacci Numbers} \label{app:NumTheor}
By setting $a=1$ and $b=0$ in equation \ref{eq:h0eigval} 
the eigenvalues can be rewritten as:
\begin{linenomath*}
\begin{equation}
\lambda_\pm = \frac{1}{2} \left [1 \pm \sqrt{5} \right ],
\end{equation}
\end{linenomath*}
so that $\lambda_+=\frac{1+\sqrt{5}}{2}$ equals the golden ratio.
Consequently, a variety of identies can be used such as:
\begin{linenomath*}
\begin{equation}
\lambda_+=\frac{-1}{\lambda_-} \ ; \ \lambda_+=1-\lambda_- \ ; \ 1+\lambda_\pm^2=\pm \lambda_+\sqrt{5}.
\end{equation}
\end{linenomath*}
Manipulating equation \ref{eq:ZLh0gf} it can be related to the closed form solution of the Fibonacci numbers:
\begin{linenomath*}
\begin{align}
Z_L(a=1,b=0) & = \frac{\lambda_+^{L+1}}{1+(1-\lambda_+)^2} + \frac{\lambda_-^{L+1}}{1+(1-\lambda_-)^2} = \Omega_{L,0} \\
    & = \frac{\lambda_+^{L+1}}{1+\lambda_-^2} + \frac{\lambda_-^{L+1}}{1+\lambda_+^2} \\
    & = \frac{\lambda_+^{L+1}}{-\lambda_- \sqrt{5}} + \frac{\lambda_-^{L+1}}{\lambda_+ \sqrt{5}} \\
    & = \frac{\lambda_+^{L+2}-\lambda_-^{L+2}}{\sqrt{5}} \ = \ F_{L+2}
\end{align}
\end{linenomath*}
where $F_n$ is the $n$th Fibonacci number.

\acknowledgments
This work was supported by the United States National Aeronautics 
and Space Administration (NASA) through grant NNX14AQ05G
 and used the used the computational resources of
 the University of Minnesota
School of Physics and Astronomy Condor cluster.
We thank Professor Gregg Musiker
in the math department for useful discussions and Ravindra Desai
for sharing his Titan data from \citet{desai2017carbon}.
The Titan data is available on NASA's Planetary 
Database System as well, in summary form in \citet{desai2017carbon}.

\listofchanges

\end{document}